\documentclass[nofootinbib,preprint,superscriptaddress]{revtex4-1}
\usepackage{amsmath,graphicx,hyperref}

\newcommand{\fms}[1]{{#1}\!\!\!/}

\newcommand{\mc}{\mathcal}
\newcommand{\mr}{\mathrm}

\newcommand{\mO}{\mathcal{O}}

\newcommand{\be}{\begin{equation}} 
\newcommand{\ee}{\end{equation}} 
\newcommand{\bea}{\begin{eqnarray}} 
\newcommand{\eea}{\end{eqnarray}}

\newcommand{\pp}{\perp}
\newcommand{\dg}{\dagger}
\newcommand{\n}{\overline{n}}
\newcommand{\nn}{\frac{\fms{\overline{n}}}{2}} 
\newcommand{\nnn}{\frac{\fms{n}}{2}} 

\newcommand{\bl}[1]{{\bf{#1}}}

\newcommand{\blp}[1]{{\bf{#1}}_{\perp}}
\newcommand{\blpu}[1]{{\bf{#1}}^{\perp}}

\newcommand{\bsp}[1]{{\boldsymbol{#1}}_{\perp}}

\newcommand{\nnb}{\nonumber} 

\newcommand{\as}{\alpha_s}

\newcommand{\veps}{\varepsilon}

\newcommand{\Lakt}{\Lambda_{\mr{k_T}}^2}

\begin{document}

%%%%%%%%%%%%%%%%%%%%%%%%%%%%%%%%%%%%%%%%%%%%%%%%%%%%%%%%%%%%%%%%%%%%%%
%%%%%%%%%%%%%%%%%%%%%%%%%%%%% Title %%%%%%%%%%%%%%%%%%%%%%%%%%%%%%%%%%
%%%%%%%%%%%%%%%%%%%%%%%%%%%%%%%%%%%%%%%%%%%%%%%%%%%%%%%%%%%%%%%%%%%%%%

\title{Factorization of Jet Mass Distribution in the small $R$ limit}

\def\Seoultech{Institute of Convergence Fundamental Studies and School of Liberal Arts, Seoul National University of Science and Technology, Seoul 01811, Korea}
\def\Duke{Department of Physics, Duke University, Durham, North Carolina 27708, USA}
\def\WS{Department of Physics, Wayne State University, Detroit, MI, 48202, USA}

\author{Ahmad Idilbi}
\email[E-mail:]{ahmad.idilbi@wayne.edu}
\affiliation{\WS} 
\author{Chul Kim}
\email[E-mail:]{chul@seoultech.ac.kr}
\affiliation{\Seoultech}

\begin{abstract} 
%\baselineskip 3.5ex  
%\vspace{0.5cm}  
We derive a factorization theorem for the jet mass distribution with a given $p_T^J$ for the inclusive production, where $p_T^J$ is a large jet transverse momentum.  Considering the small jet radius limit $(R\ll 1)$ we factorize the scattering cross section into a partonic cross section, the fragmentation function to a jet, and the jet mass distribution function. The decoupled jet mass distributions for quark and gluon jets are well-normalized and scale invariant. And they can be extracted from the ratio of two scattering cross sections such as $d\sigma/(dp_T^JdM_J^2)$ and $d\sigma/dp_T^J$.
When $M_J \sim p_T^J R$, the perturbative series expansion for the jet mass distributions works well. 
As the jet mass becomes small, the large logarithms of $M_J / (p_T^J R)$ appear, and they can be systematically resummed through more refined factorization theorem for the jet mass distribution.  
%We also obtain the resummed results of large logarithms arising from the small jet mass.    

\end{abstract}

\maketitle 

%%%%%%%%%%%%%%%%%%%%%%%%%%%%%%%%%%%%%%%%%%%%%%%%%%%%%%%%%%%%%%%%%%%%%%%%%%%%%%%%%%%%%%%
%%%%%%%%%%%%%%%%%%%%%%%%%%%%%%%%%%%%%%%%%% Content %%%%%%%%%%%%%%%%%%%%%%%%%%%%%%%%%%%%
%%%%%%%%%%%%%%%%%%%%%%%%%%%%%%%%%%%%%%%%%%%%%%%%%%%%%%%%%%%%%%%%%%%%%%%%%%%%%%%%%%%%%%%
%\baselineskip 0.1ex
\section{Introduction}
Jets, collimated bunches of hadrons, contain valuable information to study QCD and high energy interactions. 
Because jets are well localized in a certain direction, they are rather easily measurable. Also insensitiveness to long distance strong interactions enables us to handle this phenomena perturbatively. Therefore, through comparison between theoretical predictions and experiments, we are able to understand high energy interactions and explore new physics in a keen accuracy. 
%Towards this end, there have been a lot of studies of jet substructures, which improve our jet identification and characterize boosted heavy particles.  

As the energy of collisions increases,
%the produced heavy particles we are interested in can be boosted more often. So 
it is useful to employ small radius jets in order to resolve highly energetic particles into multiple jets.  
This helps us separate signals we are interested in from backgrounds. 
%such as the production of top quarks or Higgs bosons from the QCD background.
Also, with small radius jets we can suppress contaminations arising from the underlying/pile-up events. 

In a theoretical aspect, the phenomena of a jet with small raidus $R$ can be well decoupled from hard collision interactions and systematically described by collinear and (collinear-)soft interactions. Also we can effectively ignore or detach soft gluon emissions with a wide angle from the jet direction. 
However, the small radius induces large logarithms in the perturbative calculation of $\as$. Hence we have to resum the large logarithms to all orders for reliable predictions~\cite{Dasgupta:2014yra,Dasgupta:2016bnd,Kang:2016mcy,Dai:2016hzf}. 

Jet mass is one of the most important jet substructures. 
When a boosted heavy particle such as a top or Higgs boson decays into a jet,   
we can identify the heavy particle from the peak of the jet mass if we can separate QCD jets effectively.
Therefore a precise description of QCD jet mass distribution is prerequisite for the identification of the heavy particle and moreover understanding physics at TeV scales. 

So far the QCD jet mass distributions~\cite{Banfi:2010pa,Dasgupta:2012hg,Chien:2012ur,Jouttenus:2013hs,Kolodrubetz:2016dzb} have been widely studied focusing on resumming large logarithms of the small jet mass compared to large $E_J$ or $p_T^J$, where $E_J$ is the jet energy and $p_T^J $ is the jet transverse momentum with respect to an incoming beam axis.
In the small $R$ limit a typical scale for the jet mass is comparable to $E_JR$ or $p_T^J R$. So, for example, if $p_T^J$ is a few TeV, we can describe the jet mass up to a few hundreds of GeV employing the small $R$ approximation.\footnote{\baselineskip 3.0ex For practical use, the size of $R$ needs not to be too small. 
It is known that the small $R$ approximation practically works well even for the case when $R$ is 0.6 - 1~\cite{Jager:2004jh,Dasgupta:2012hg}.}  
%Therefore, for example, if we consider a top with $p_T \sim \mO(\mr{TeV})$, most of top decayed products can be recombined to a top jet even with small $R$. 
However, in the perturbative expansion for QCD jet, the distribution of the nonzero jet mass is 
%proportional to $\delta(M_J)$ at leading order (LO) in $\as$ and
roughly given as $\as/M_J$. Thus the region $M_J\ll p_T^JR$ is dominant and the resummation of the large logarithms such as $\ln (p_T^J R/M_J)$~\cite{Kolodrubetz:2016dzb} is inevitable. 
In this case we have to explore QCD dynamics setting the scale hierarchy as $(E_J,~p_T^J) \gg (E_JR,~p_T^JR) \gg M_J$. 

In this paper, for the inclusive jet production, we introduce the normalized jet mass distribution, which can be universally applied to any isolated QCD jet with small $R$. With the help of the fragmentation functions to a jet (FFJs)~\cite{Kang:2016mcy,Dai:2016hzf} we show that the distribution can be separated from the jet scattering cross section with a given $p_T^J$ or $E_J$ up to next-to-leading order (NLO) in $\as$. The jet mass distribution turns out to be scale invariant and can describe the peak region$~(M_J\ll p_T^J R)$ as well as the tail region$~(M_J\sim p_T^JR)$. We also derive a more refined factorization theorem in focusing on the peak region. This allows us to handle large logarithms arising from the small jet mass compared to $p_T^JR$. 

\section{Factorization theorem for the jet mass distribution}

\subsection{Collinear Factorization for Jet Substructure}

When we consider the inclusive jet scattering cross section such as $N_1N_2\to JX$ and the observed jet $J$ has a small radius $R ~(\ll 1)$, we can factorize the scattering cross section as~\cite{Dasgupta:2014yra}
\be
\label{infacj}
\frac{d\sigma}{dydp_T^J} = \sum_i\int^1_{z_J=p_T^J/Q_T} \frac{dz}{z} \frac{d\sigma_i(y,z_J/z,\mu)}{dy dp_T^i} D_{J/i} (z,\mu).
\ee
Here $\sigma_i$ is the cross section with a parton $i$ in the final state and $Q_T$ is the maximal $p_T$ for a given rapidity $y$. Since we consider a jet in the central rapidity region, the rapidity $y$ is given to be $\lesssim \mO(1)$. $D_{J/i}$ is the so-called fragmentation function to a jet (FFJ)~\cite{Kang:2016mcy,Dai:2016hzf} for a given mother parton $i$. The FFJ is described as the probability that the outgoing jet from the mother parton acquires a large momentum fraction $z$. 

For detailed perturbative results for a jet, throughout this paper we consider anti-$\mr k_{\mr{T}}$ algorithm~\cite{Cacciari:2008gp} for the clustering. 
At NLO in $\as$, for $\mr k_{\mr{T}}$-type algorithms to include $\mr k_{\mr{T}}$~\cite{Catani:1993hr,Ellis:1993tq}, anti-$\mr k_{\mr{T}}$, and Cambridge/Aachen (C/A)~\cite{Dokshitzer:1997in}, the jet merging conditions of two particle emissions are given as 
\be \label{jmc}
\theta < R'~ 
\left\{
\begin{array}{rl} 
& R'= R ~~\mbox{for $e^+e^-$ collider} \\ 
&R'=R/\cosh y  ~~\mbox{for hadron collider.} \end{array} \right.
\ee
Here $\theta$ is the angle between the two particles, and $\Delta R = \sqrt{\Delta y^2 +\Delta \phi^2}$ for the hadron collider is assumed to be small and can be approximated by $\Delta R \sim \theta \cosh y$. 
%Our result for $\Phi_{J/i}$, using Eq.~(\ref{jmc}), are applicable to $e^+e^-$ and hadron colliders as well. 
In our computation as we will see, the typical scale for the observed jet is given as $p_J^+ \tan(R'/2) \sim E_J R'$. This is expressed as $E_JR$ for $e^+e^-$ annihilation and $p_T^J R$ for hadronic collision.

Based on Eq.~(\ref{infacj}), we can also investigate substructures of the observed jet. For example, if we try to figure out the fragmentation to a hadron or subjet inside the jet we can employ the following factorization theorem~\cite{Dai:2016hzf}:
\be
\label{infacsj}
\frac{d\sigma}{dydp_T^Jdx} = \sum_{i,k}\int^1_{z_J} \frac{dz}{z} \frac{d\sigma_i(y,z_J/z,\mu)}{dy dp_T^i} D_{J_k/i} (z;E_iR',\mu) D_{l/J_k}(x;E_JR'),
\ee
where $l$ is the hadron or subjet inside $J_k$, $k$ denotes a primary parton flavor for the observed jet, and $x=p_T^l/p_T^J$ is a momentum fraction of $l$ over $J_k$. $D_{l/J_k}$ is the so-called jet fragmentation function (JFF), which describes a fragmenting process $k\to l$ inside $J_k$.

In Eq.~(\ref{infacsj}), the FFJs and JFFs are normalized as one and satisfy the momentum conserving sum rule such as 
\be
\label{sumr}
1=\sum_{k} \int^1_0 dz z D_{J_k/i}(z;E_iR',\mu),~~~1=\sum_{l} \int^1_0 dz z D_{l/J_k}(z,E_JR').
\ee
Note that the JFFs, $D_{l/J_k}$, have a limited phase space since whole partons should radiate only inside the jet. To describe the fragmenting process using soft-collinear effective theory (SCET)~\cite{Bauer:2000ew,Bauer:2000yr,Bauer:2001yt,Bauer:2002nz}, we introduce the so-called unnormalized JFF such as  
\bea 
\tilde{D}_{l/J_q}(z,\mu) = \sum_{X \in J}\frac{1}{2N_cz}  \int d^{D-2}\blpu{p} &&  \mr{Tr} \langle 0 | \delta \Bigl(\frac{p_+}{z}-\mc{P}_+\Bigr)\delta^{(D-2)}(\bsp{\mc{P}})\nn \Psi_n | l(p_+, \blpu{p}) X\rangle \nnb\\ 
\label{nJFF}
~~~~~~&&\times\langle l(p_+, \blpu{p}) X| \bar{\Psi}_n |0\rangle.
\eea
Here $D=4-2\veps$, and the collinear quark is given by $\Psi_n=W_n^{\dagger} \xi_n$, where $W_n$ is a collinear Wilson line in SCET~\cite{Bauer:2000yr,Bauer:2001yt}.
If we consider a gluon jet as an initial state, the fragmentation $\tilde{D}_{l/J_g}$ can be similarly described by a collinear gluon field strength $\mc{B}_{n}^{\pp a} = i\n^{\rho}g_{\perp}^{\mu\nu} G_{n,\rho\nu}^b \mc{W}_n^{ba} = i\n^{\rho}g_{\perp}^{\mu\nu} \mc{W}_n^{\dagger,ba} G_{n,\rho\nu}^b$, where $\mc{W}_n$ is the collinear Wilson line in the adjoint representation.  
We decomposed the momentum as $p^{\mu} = (p_+,p_-,\blp{p})$, where $p_+\equiv \n\cdot p = p_0 + \hat{\bl{n}}_J\cdot \bl{p}$ and $p_-\equiv n\cdot p = p_0 -\hat{\bl{n}}_J\cdot \bl{p}$. Here $\hat{\bl{n}}_J$ is an unit vector in the jet direction and $\blp{p}$ is a transverse momentum to $\hat{\bl{n}_J}$. The lightcone vectors $n$ and $\n$ satisfy $n^2=\n^2=0$ and $n\cdot\n =2$. 

The definition of the unnormalized JFF in Eq.~(\ref{nJFF}) are almost the same as the usual fragmentation function (FF). The only difference is that the final states for the unnormalized JFF should be inside the jet with a size $E_JR'$, while the usual FF has not such a restriction. So, as computed in Refs.~\cite{Procura:2011aq,Dai:2016hzf} at NLO, the normalization of $\tilde{D}_{l/J_k}$ differs from one due to a limited phase space and it is given as a function of $E_JR'$ such as  
\be
\label{normnJFF}
\sum_{l} \int^1_0 dz z \tilde{D}_{l/J_k}(z;E_JR',\mu) = \mc{J}_k(E_JR',\mu).
\ee
We call it as `the integrated jet function (inside a jet)', which describes parton radiations inside a jet. 
Therefore the normalized JFF shown in Eqs.~(\ref{infacsj}) and (\ref{sumr}) is obtained from dividing the unnormalized JFF by the integrated jet function such as 
\be
D_{l/J_k}(z;E_JR') = \frac{\tilde{D}_{l/J_k}(z;E_JR',\mu)}{\mc{J}_k((E_JR',\mu)}.
\ee

This integrated jet function is also needed for describing the `in-jet' contributions to the FFJs, which have the following structure: 
\be\label{FFJB} 
D_{J_k/i} (z,\mu; ER') =  B_{J_k/i} (z;ER',\mu) \mc{J}_k(\mu;E_JR'), 
\ee 
where the jet splitting kernel $B_{J_k/i}$ can be expressed as 
\be\label{Bjki} 
B_{J_k/i} (z;ER',\mu) = \delta(1-z)\delta_{ik} +D^{\rm{out}}_{J_k/i}(z;ER',\mu).
\ee
Here $D^{\rm{out}}_{J_k/i}(z)$ is the jet splitting (`out-jet') contributions to the FFJs. This factorization for the FFJs in Eq.~(\ref{FFJB}) works at NLO in $\as$. It might hold at the higher order if we ignore $1\to3$ splitting processes. 
In deriving Eq.~(\ref{infacsj}) we can use the fact that the fragmentation from $k$ to $l$ can be factorized as the out-jet and in-jet splitting processes such as 
\be\label{FFfac} 
D_{l/k} (x) = \sum_k \int^1_w \frac{dz}{z} B_{J_k/i}\left(z;ER'\right) \tilde{D}_{l/J_k} \left(\frac{x}{z};E_J R'\right).
\ee 
Then we multiply/divide the integrated jet function, and finally obtain the factorization theorem in Eq.~(\ref{infacsj}).

In Eq.~(\ref{infacsj}) 
if the momentum fractions $z$ and $x$ are $\sim\mc{O}(1)$ and not too close to 1, we can genuinely describe the FFJs and JFFs by a collinear mode. And we can successfully suppress the contributions from (collinear-)soft degrees of freedom\footnote{\baselineskip 3.0ex
Here the soft degrees of freedom can be separated as a regular soft mode and a collinear-soft mode. The former scales as $p_s \sim Q\eta(1,1,1)$ and the latter as $p_{cs} \sim Q\eta(1,\lambda^2,\lambda)$, where $\eta$ and $\lambda$ are relevant small parameters to physical situations. The regular soft mode does not contribute to the computation of the jet in the small $R$ limit since it cannot recognize the jet boundary and describe radiations only far outside the jet. The collinear-soft mode~\cite{Bauer:2011uc,Procura:2014cba,Becher:2015hka,Chien:2015cka} describes radiations of boosted soft particles near the jet boundary and can contribute if a jet observable is sensitive to a small momentum. For a detailed decoupling procedure of the collinear-soft mode from the collinear mode we refer to Refs.~\cite{Bauer:2011uc,Dai:2017dpc}.
}, 
which can be decoupled from the collinear mode. 
Here the collinear mode scales as $p_{c} = (p_{c}^+,p_{c}^-,\blpu{p}_{c}) \sim Q(1,R^2,R)$, 
where $Q \sim p_T^J\sim E_J$.
Hence it recognizes the jet boundary and gives nonvanishing results for both in-jet and out-jet contributions.
If $z$ or $x$ in Eq.~(\ref{infacsj}) are close to 1, the collinear-soft contributions do not vanish and the decoupled collinear-soft mode is responsible for radiations with momentum $Q(1-z)$ or $Q(1-x)$~\cite{Dai:2017dpc}. 

\subsection{Factorization to Jet Mass Distribution in the Tail Region: $M_J \sim E_J R'$}

From now we consider the factorization into a jet mass distribution starting from Eq.~(\ref{infacsj}). In the tail region of the jet mass distribution, the jet mass $M_J$ is comparable with the jet size $E_JR'$. In this case the collinear mode with $p_{c}\sim Q(1,R^2,R)$ is enough for describing the jet mass distribution, and the collinear-soft contributions are suppressed. In order to incorporate the jet mass distribution with Eq.~(\ref{infacsj}), we introduce fragmenting jet function (FJF) inside a jet~\cite{Procura:2011aq} putting $\delta(M_J^2-\mc{P}^2)$ into the collinear operator for the JFF such as
\bea 
\mc{G}_{l/J_q}(x,M_J^2;E_JR',\mu) = \sum_{X \in J}\frac{1}{2N_cx}  \int d^{D-2}\blpu{p} &&  \mr{Tr} \langle 0 | \delta \Bigl(\frac{p_+}{x}-\mc{P}_+\Bigr)\delta^{(D-2)}(\bsp{\mc{P}})\delta(M_J^2-\mc{P}^2)\nn \Psi_n  \nnb\\ 
\label{FJF}
~~~~~~&&\times| l(p_+, \blpu{p}) X\rangle\langle l(p_+, \blpu{p}) X| \bar{\Psi}_n |0\rangle.
\eea
At leading order (LO) in $\as$, the FJF is normalized as $\mc{G}_{l/J_q}^{(0)} = \delta(1-x) \delta(M_J^2)$. The difference compared to a generic FJF~\cite{Procura:2009vm,Jain:2011xz,Ritzmann:2014mka} is that the upper limit of the jet mass is constrained by a jet algorithm, while the generic FJF does not have such a constraint. 
Then we have the relation 
\be
\label{FJFrel}
D_{l/J_k}(x;E_JR') = \frac{1}{\mc{J}_k (E_JR',\mu)}\int_0^{\Lambda^2(x)} dM_J^2~ \mc{G}_{l/J_k}(x,M_J^2;E_JR',\mu),   
\ee
where $\Lambda^2(x)$ is the maximal jet mass for a given $x$. In case of $\mr{k_T}$-type algorithms $\Lambda^2(x)$ is given by $x(1-x) p_J^{+2} t^2$, where $p_J^+ \sim 2E_J$ and $t \equiv \tan(R'/2) \sim R'/2$. 

Thus, from Eq.~(\ref{FJFrel}), the differential cross section to include information on the jet mass can be written as 
\be\label{xsec0} 
\frac{d\sigma}{dydp_T^JdxdM_J^2} = \int^1_{z_J} \frac{dz}{z} \frac{d\sigma_i(y,z_J/z,\mu)}{dy dp_T^i}  D_{J_k/i} (z;ER',\mu) \frac{\mc{G}_{l/J_k}(x,M_J^2;E_JR',\mu)}{\mc{J}_k (E_JR',\mu)}. 
\ee 
If we apply momentum sum rule over the final states $l$, we also obtain 
\bea
\frac{d\sigma}{dydp_T^JdM_J^2} &=& \sum_l \int_0^1 dx x \left(\frac{d\sigma}{dydp_T^JdxdM_J^2}\right) \nnb \\
\label{xsecm}
&=& \int^1_{z_J} \frac{dz}{z} \frac{d\sigma_i(y,z_J/z,\mu)}{dy dp_T^i}  D_{J_k/i} (z;ER',\mu) \Phi_k (M_J^2;E_JR'). 
\eea
Here $\Phi_k$ is our desired jet mass distribution for a given $E_J$ and $R'$, and written as  
\be
\label{jmd} 
\Phi_k (M_J^2;E_JR') = \sum_l \int_0^1 dx x \frac{\mc{G}_{l/J_k}(x,M_J^2;E_JR',\mu)}{\mc{J}_k (E_JR',\mu)}. 
\ee

As seen from Eqs.~(\ref{FJFrel}) and (\ref{xsecm}), $\Phi_k$ is scale invariant except the dependence of $\as(\mu)$ in the perturbative series. In Eq.~(\ref{xsecm}), the convolution of $d\sigma_i$ and the FFJs is already given to be scale invariant since the renormalization behavior of the FFJs follow DGLAP evolutions resumming large logarithms of small $R$~\cite{Dasgupta:2014yra,Dasgupta:2016bnd,Kang:2016mcy,Dai:2016hzf}. Also $\Phi_k$ is well-normalized to satisfy $1 = \int dM_J^2 ~\Phi_k(M_J^2)$. This fact can be clearly seen if we apply the momentum sum rule to Eq.~(\ref{FJFrel}). Since $\Phi_k$ is decoupled from the hard scattering process (and the FFJs) as shown in Eq.~(\ref{xsecm}), it can be universally determined for a given jet with $p_T^J$ and $R$, and can be also applied to $e^+e^-$ annihilation and deep inelastic scattering.

%We will describe $\Phi_{J/i}$ in Eq.~(\ref{xsec0}) in the framework of soft-collinear effective theory (SCET)~\cite{Bauer:2000yr,Bauer:2001yt}, where collinear and soft modes are separated in a gauge invariant way. Collinear mode in the jet direction for $\Phi_{J/i}$ scales as $p_{n_J}=(p_+,p_{\pp},p_-)\sim Q(1,R,R^2)$, where $Q \sim p_T^J$ is a hard scale and $R$ is comparable to the small parameter $\lambda$ in SCET. In our convention $p_{\pm}$ are denoted as $p_+\equiv \n_J\cdot p = p^0+ \hat{\bl{n}}_J \cdot \bl{p}$ and $p_-\equiv n_J\cdot p = p^0 - \hat{\bl{n}}_J \cdot \bl{p}$, where $\hat{\bl{n}}_J$ is an unit vector in the jet direction. 

At NLO in $\as$, with the clustering condition in Eq.~(\ref{jmc}) applied, the normalization factor in Eq.~(\ref{jmd}), i.e, the integrated jet functions $\mc{J}_k$, are computed  as~\cite{Cheung:2009sg,Ellis:2010rwa,Chay:2015ila}
\bea\label{qintj} 
\mc{J}_q(E_J R',\mu) &=& 1+\frac{\as C_F}{2\pi} \Biggl[%\frac{1}{\UV^2}+ \frac1{\UV}\Bigl(\frac32 + \ln\frac{\mu^2}{p_J^{+2} t^2}\Bigr)
%\nnb\\
%&&~~~~~~~~~~~
 \frac{3}{2}\ln\frac{\mu^2}{p_{J}^{+2}t^2}+\frac{1}{2}\ln^2\frac{\mu^2}{p_{J}^{+2}t^2}+\frac{13}{2}-\frac{3\pi^2}{4} \Biggr]\ , 
\\
\mc{J}_g(E_J R',\mu) &=& 1+\frac{\as C_A}{2\pi} \Biggl[%\frac{1}{\UV^2}+\frac{1}{\UV}\Bigl(\frac{\beta_0}{2C_A} +\ln\frac{\mu^2}{p_{J}^{+2}t^2}\Bigr) +
\frac{\beta_0}{2C_A}\ln\frac{\mu^2}{p_{J}^{+2}t^2}%\nnb \\ 
\label{gintj}
%&&~~~~~~~~~~~
+\frac{1}{2}\ln^2\frac{\mu^2}{p_{J}^{+2}t^2}+\frac{67}{9}-\frac{23n_f}{18C_A}-\frac{3\pi^2}{4} \Biggr]\ ,
\eea
where $p_J^+ t \sim E_J R'$, $\beta_0 = 11N_c/3-2n_f/3$, $C_A=N_c=3$, and $n_f$ is the number of quark flavors.

Computing the FJFs $\mc{G}_{l/J_k}$ in Eq.~(\ref{jmd}) and dividing Eqs.~(\ref{qintj}) and (\ref{gintj}), we obtain the normalized jet mass distributions at NLO such as 
\bea 
\label{Phiq}
\Phi_q(M_J^2) &=& \delta(M_J^2)+\frac{\as C_F}{2\pi} \Biggl[ \frac{1}{M_J^2} \Bigl( -\frac{3}{2}w  +2 \ln \frac{1+w}{1-w}\Bigr)\Biggr]_{\Lambda^2=\Lakt,} \\
\label{Phig}
\Phi_g(M_J^2) &=& \delta(M_J^2)+\frac{\as C_A}{2\pi}\Biggl[\frac{1}{M_J^2}\Bigl(-\frac{7w}{4}-\frac{w^3}{12}+2 \ln \frac{1+w}{1-w}\Bigr) +\frac{n_f}{C_A M_J^2}\Bigl(\frac{w}{4}+\frac{w^3}{12}\Bigr)\Biggr]_{\Lambda^2=\Lakt,}
\eea 
where $w=\sqrt{1-M_J^2/\Lakt}$, and $\Lakt= p_J^{+2}t^2/4 \sim E_J^2R'^2/4$ is the maximum of the jet mass with anti-$\mr{k_T}$ algorithm applied.
$[\cdots]_{\Lambda^2}$ is the  so-called ``$\Lambda^2$-distribution'', which is defined as 
\be\label{Ldist}
\int_0^{\mc{M}^2} dM^2[g(M^2)]_{\Lambda^2} f(M^2)  = 
\int_0^{\mc{M}^2}dM^2 g(M^2) f(M^2) -\int_0^{\Lambda^2} dM^2 g(M^2) f(0),
\ee
where $f(M^2)$ is an arbitrary smooth function at $M^2=0$. 
In computing Eqs.~(\ref{Phiq}) and (\ref{Phig}), the one loop contributions to $\delta(M_J^2)$ for $\sum_l \int dx x\mc{G}_{l/J_k}$ are cancelled by $\mc{J}_k$ at one loop. Then we obtain the scale invariant jet mass distributions, that is also  normalized as one.\footnote{\baselineskip 3.0 ex
In computing $\sum_l \int dx x\mc{G}_{l/J_k}$ to obtain the jet mass distributions, we applied the zero-bin subtraction~\cite{Manohar:2006nz} to eliminate the overlap with soft contributions. Here the subtracted contribution does not come from the regular soft mode but the collinear-soft mode scaling as $\sim Q\eta(1,R^2,R)$. As shown in Ref.~\cite{Chien:2015cka,Chay:2015ila}, the different zero-bin subtraction by the collinear-soft mode gives a different renormalization behavior from the standard jet function that was defined in Eq.~(\ref{indJ}).}

As far as $M_J \sim E_JR'$, there is no large logarithm in Eqs.~(\ref{Phiq}) and (\ref{Phig}). However, as $M_J$ goes to zero, the large logarithm, $\ln M_J^2/(E_J^2R'^2)$, appears. In order to see thee small jet mass behaviors of $\Phi_k$, we need to use a small value of $\Lambda^2$ for the $\Lambda^2$-distribution instead of using $\Lambda^2=\Lakt$.
%for an upper value (in the subtraction term in Eq.~(\ref{Ldist})). 
From Eq.~(\ref{Ldist}), using the relation 
\be
[g(M^2)]_{\Lakt} = [g(M^2)]_{\Lambda^2} - \delta(M^2) \int^{\Lakt}_{\Lambda^2} dM'^2 g(M'^2), 
\ee
we can rewrite $\Phi_{k=q,g}$ in Eqs.~(\ref{Phiq}) and (\ref{Phig}). Here $\Lambda^2$ in the right side scales as $\Lambda^2 \ll E_J^2R'^2$. Then we can safely take a limit $M_J^2 \to 0$ on $\Phi_k$, and obtain 
\bea
\Phi_q(M_J^2\to 0) &=& \delta(M_J^2)+\frac{\as C_F}{2\pi}\Biggl\{\delta(M_J^2)\Bigl(-\frac{3}{2} \ln\frac{\Lambda^2}{p_J^{+2}t^2}-\ln^2\frac{\Lambda^2}{p_J^{+2}t^2}-3+\frac{\pi^2}{3}\Bigr) \nnb\\
\label{Phiqsm}
&&\hspace{3cm}-\Biggl[\frac{1}{M_J^2} \Bigl( \frac{3}{2}  +2\ln\frac{M_J^2}{p_J^{+2}t^2}\Bigr)\Biggr]_{\Lambda^2\ll E_J^2R'^2,} \Biggr\}_,\\
\Phi_g(M_J^2\to 0) &=& \delta(M_J^2)+\frac{\as C_A}{2\pi}\Biggl\{\delta(M_J^2)\Bigl(-\frac{\beta_0}{2C_A} \ln\frac{\Lambda^2}{p_J^{+2}t^2}-\ln^2\frac{\Lambda^2}{p_J^{+2}t^2}-\frac{67}{18}+\frac{13}{18}\frac{n_f}{C_A}+\frac{\pi^2}{3}\Bigr) \nnb\\
\label{Phigsm}
&&\hspace{3cm}-\Biggl[\frac{1}{M_J^2} \Bigl( \frac{\beta_0}{2C_A}  +2\ln\frac{M_J^2}{p_J^{+2}t^2}\Bigr)\Biggr]_{\Lambda^2\ll E_J^2R'^2,} \Biggr\}_.
\eea
So we clearly see the large logarithms of $M_J^2/(E_J^2R'^2)$ in the small jet mass limit, that needs to be resummed to all order in $\as$. For this we need more IR sensitive modes to be decoupled from the collinear mode. In the next section, including these modes we will consider the factorization theorem for the small jet mass distributions. 

\section{Factorization of the Jet Mass Distribution in the Peak Region: $M_J^2 \ll E_J^2 R'^2$}

If we assume $M_J^2 \ll E_J^2R'^2$, the collinear mode with $p_{c}\sim Q(1,R^2,R)$ cannot radiate inside a jet since its contribution to the jet mass squared is given to be $\sim Q^2R^2$. Therefore the collinear contributions are involved only in the normalization factor $\mc{J}_k$ when we consider the distribution $\Phi_k$. Then the FJF $\mc{G}_{l/J_k}$ in Eq.~(\ref{jmd}) should be described by the narrower collinear mode. We will call it `ultracollinear mode', and its scaling behavior is given as $p_{uc} \sim Q(1,M_J^2/Q^2,M_J/Q)$ satisfying $p_{uc}^2 \sim M_J^2$. Since this mode is too narrow to be aware of the jet boundary, the FJF can be identified as a generic FJF without the constraint of the boundary. 

Hence if we apply the momentum sum rule to $\mc{G}_{l/J_k}$ in the ultracollinear limit, it ends up as `the standard jet function' such as~\cite{Procura:2009vm,Jain:2011xz}
\be
\label{indJ}
J_k (M_c^2,\mu) = \sum_{l} \int_0^1 dx x~\mc{G}_{l/J_k}(x,M_c^2;\mu), 
\ee
where $M_c$ is the ultracollinear contribution to the jet mass, and we suppressed the term $E_JR'$ in the argument of $\mc{G}_{l/J_k}$ since it does not appear in the ultracollinear limit. $J_k$ is the standard jet function that was first introduced in Ref.~\cite{Bauer:2001yt}. And, for example, the quark jet function is defined as  
\be
\sum_{X_n} \langle 0 | \Psi_n^{\alpha} |X_n \rangle \langle X_n | \bar{\Psi}_n^{\beta} | 0 \rangle \\
\label{jetfq}
=\int \frac{d^4 p}{(2\pi)^3} p_+ \nnn J_{q}(p^2,\mu) \delta^{\alpha\beta},  
\ee
where the jet function at LO is normalized as $J_q^{(0)} = \delta (p^2)$. 

Because the jet mass is small, it can be also sensitive to collinear-soft radiations. 
As discussed before, the decoupled collinear-soft mode from the collinear mode does not contribute when $M_J \sim E_JR'$. However, if $M_J \ll E_JR'$, the momentum squared of the ultracollinear and the collinear-soft modes can be comparable to the small jet mass such as $p_{uc}^+p_{cs}^- \sim p_J^+ p_{cs}^- \sim M_J^2$ . In general the scaling behavior of the collinear-soft momentum can be written as $p_{cs} \sim Q\eta(1,\lambda^2,\lambda)$, where $\eta$ and $\lambda$ are small parameters. Here $\lambda$ is given by $\lesssim R$ in order that the collinear-soft mode contribute to the jet mass inside the jet. 
Also using the fact $p_J^+ p_{cs}^- \sim M_J^2$, we estimate $\eta \sim M_J^2/(Q^2 \lambda^2) \ll 1$. Since we are now interested in the region $M_J^2 \ll Q^2R^2$, $\lambda$ should not be much less than $R$, otherwise the small parameter $\eta$ could be $\mc{O}(1)$. As a result the small parameter $\lambda$ is given as $\mc{O}(R)$, and finally the collinear-soft momentum scales as 
\be\label{spcs}
p_{cs}^{\mu} = (p_{cs}^+,p_{cs}^-,\blpu{p}_{cs}) \sim \frac{M_J^2}{QR^2} (1,R^2,R).
\ee 
Note that this collinear-soft mode can read the jet boundary properly like the collinear mode $p_c \sim Q(1,R^2,R)$.

From the collinear mode, we can decouple the collinear-soft interactions following the similar procedure performed in Ref.~\cite{Dai:2017dpc}. Then the decoupled collinear-soft interactions can be expressed in terms of the collinear-soft Wilson lines $Y_{n,cs}$ and $Y_{\n,cs}$, which have the usual form of the soft Wilson lines~\cite{Bauer:2001yt,Chay:2004zn} such as 
\be\label{csWil} 
Y_n^{cs}(x) = \mr{P}\exp \Biggl[ig\int^{\infty}_x ds n\cdot A_{cs} (sn)\Biggr]\ ,~~~
Y_{\n}^{cs}(x) = \mr{P}\exp \Biggl[ig\int^{\infty}_x ds \n\cdot A_{cs} (s\n)\Biggr]\ . 
\ee

Finally, incorporating the collinear-soft interactions with Eq.~(\ref{jmd}) and applying Eq.~(\ref{indJ}), we obtain the factorized result of the jet mass distribution for the region $M_J\ll E_JR'$ such that 
\bea 
\Phi_{k} (M_J^2\ll E_J^2R'^2;E_JR') &=& \mc{C}_k(E_JR',\mu)\int dM_c^2 d\ell_- \delta (M_J^2 - M_c^2 - p_J^+ \ell_-) \nnb \\
&&~~~~~~~~~~~~~\times J_{k} (M_c^2;\mu) \tilde{S}_{k} (\ell_-;E_JR',\mu) \nnb \\
\label{facjmd}
&=&\mc{C}_k(E_JR',\mu)\int_0^{M_J^2} dM_c^2 J_{k} (M_c^2;\mu) {S}_{k} (M_J^2-M_c^2;E_JR',\mu),
\eea 
where the collinear function $\mc{C}_k $ is equal to $\mc{J}_k^{-1}$, and $J_k$ are the standard jet function introduced in Eq.~(\ref{indJ}). 
$\tilde{S}_k (\ell_-)$ are the collinear-soft functions, and for $k=q$ it is defined as 
\be\label{csoftF} 
\tilde{S}_q (\ell_-)  = \frac{1}{N_c} \mr{Tr}~\langle 0 | Y_{\n,cs}^{\dg} Y_{n,cs} \delta(\ell_-+\Theta(R'-\theta) \mc{P}_s^- ) Y_{n,cs}^{\dg} Y_{\n,cs} |0\rangle_, 
\ee
where $\mc{P}_{cs}^-$ is the derivative operator extracting the momentum $p_{cs}^-$. The collinear-soft function initiated by a gluon jet can be also defined similarly in the adjoint representation with a normalization factor $N_c^2-1$. At leading order in $\as$, $\tilde{S}_k$ is normalized as $\delta(\ell_-)$. From the argument of the delta function in Eq.~(\ref{csoftF}), we see that, at the higher order in $\as$, $\ell_-$ returns nonzero value only from the in-jet contributions, while the out-jet contributions are proportional to $\delta(\ell_-)$. $S_k(M_s^2 = M_J^2-M_c^2)$ in the last line of Eq.~(\ref{facjmd}) has the dimension $1/M^2$, and it can be related as $S_k(M_s^2) = \tilde{S}_k(\ell_-)/p_J^+$ with $M_s^2 = p_J^+ \ell_-$. 

For one collinear-soft gluon emission inside a jet, the phase space constraint for the collinear-soft momentum $k$ is given by 
\be
R'>\theta~~\to~~
\tan^2 \frac{R'}{2} > \frac{k_-}{k_+}\ .
\ee
Employing this we computed the collinear-soft function at NLO and the results are shown as 
\be 
S_k (M^2) = \delta(M^2) + \frac{\as C_k}{\pi} \Biggl\{\delta(M^2) \left(\frac{\pi^2}{24}-\frac{1}{4}\ln^2\frac{\mu^2p_J^{+2}t^2}{\Lambda^2}\right) 
\label{csoftnlo}
+\Biggl[\frac{1}{M^2} \ln\frac{\mu^2p_J^{+2}t^2}{(M^2)^2}
\Biggr]_{\Lambda^2}\Biggr\}_,
\ee
where $C_k$ are $C_F$ for $k=q$ and $C_A$ for $k=g$. The results are infrared (IR) finite, and the part with $M^2\neq 0$ has been expressed through the $\Lambda^2$-distribution defined in Eq.~(\ref{Ldist}). 
%When compared with Eqs.~(\ref{Phiq}) and (\ref{Phig}), 
The value of $\Lambda^2$ in the distribution of Eq.~(\ref{csoftnlo}) can be arbitrary since the dependence of $\Lambda^2$ in the distribution can cancel when combined with the part proportional to $\delta(M^2)$. But, the scaling behavior here can be  considered as $\Lambda^2 \sim M_J^2~(\ll E_J^2 R'^2)$. From Eq.~(\ref{csoftnlo}) we read the collinear-soft scale to minimize the large logarithms as $\mu\sim \mu_{cs} \sim M_J^2/(E_JR')$. This coincides with the fact $\mu_{cs}^2 \sim p_{cs}^2$. 

%The upper limit for $M^2$ is arbitrarily assigned as $\Lambda^2$ which is power counted as $\mO(\lambda^4)$.Since $t\equiv \tan(R'/2) \sim \mO(\lambda)$, the scale to minimize the large logarithms should is: $\mu \sim \mO(Q\lambda^3)$. 

In the factorization theorem in Eq.~(\ref{facjmd}), the standard jet functions for ultracollinear interactions are expressed at one loop such as 
\bea
\label{nloJq}
J_q (M^2) &=& \delta(M^2)  + \frac{\as C_F}{2\pi} \Biggl\{\delta(M^2) \left(\frac{3}{2}\ln\frac{\mu^2}{\Lambda^2}
+\ln^2\frac{\mu^2}{\Lambda^2}+\frac{7}{2}-\frac{\pi^2}{2}\right) \\
&&\hspace{3cm}
-\Biggl[\frac{1}{M^2}\Bigl(\frac{3}{2}+ 2\ln\frac{\mu^2}{M^2}\Bigr)\Biggr]_{\Lambda^2} \Biggr\}_,\nnb
 \\
\label{nloJg}
J_g (M^2) &=& \delta(M^2)  + \frac{\as C_A}{2\pi} \Biggl\{\delta(M^2) \left(\frac{\beta_0}{2C_A}\ln\frac{\mu^2}{\Lambda^2}
+\ln^2\frac{\mu^2}{\Lambda^2}+\frac{67}{18}-\frac{5}{9}\frac{n_f}{C_A}-\frac{\pi^2}{2}\right)\\
&&\hspace{3cm}
-\Biggl[\frac{1}{M^2} \Bigl(\frac{\beta_0}{2C_A}+2\ln\frac{\mu^2}{M^2}\Bigr)\Biggr]_{\Lambda^2} \Biggr\}_.\nnb
\eea 

Now we have obtained all ingredients for NLO computation of $\Phi_k (M_J^2\ll E_J^2 R'^2)$ in the framework of factorization. Putting NLO results for $\mc{C}_k(\mc{J}_k^{-1})$ (Eqs.~(\ref{qintj}) and (\ref{gintj})), $S_{k}$ (Eq.~(\ref{csoftnlo})), and $J_k$ (Eqs.~(\ref{nloJq}) and (\ref{nloJg})) into Eq.~(\ref{facjmd}), we easily reproduce the results in Eqs.~(\ref{Phiqsm}) and (\ref{Phigsm}), that are the asymptotic distributions when the small jet mass limit is taken into account in the calculation with the collinear mode.\footnote{\baselineskip 3.0 ex
Since the ultracollinear and collinear-soft modes can be considered as subsets of the collinear mode, we can regard the collinear mode as a `full mode' in some sense. So the jet mass distribution with the collinear mode for $M_J \sim E_JR'$ can cover the full range of the jet mass although we need the resummation of the large logarithms of $M_J/E_JR'$ in the small jet mass region. 
}

\section{Resummation for the jet mass distribution in the small jet mass region} 

As we have seen from Eqs.~(\ref{Phiqsm}) and (\ref{Phigsm}), the results for the small jet mass at the fixed order in $\as$ is not enough for the precise estimation since the perturbative expansion is not reliable due to the large logarithms. So using the factorization theorem established in Eq.~(\ref{facjmd}) we have to resum the large logarithms to all order in $\as$.    

The factorized parts $\mc{C}_k$, $J_k$, and $S_k$ in Eq.~(\ref{facjmd}) are governed by the collinear, ultracollinear, and collinear-soft modes respectively. So the relevant scale to each factorized part is given as $\mu_c \sim E_JR'$, $\mu_{uc} \sim M_J$, and $\mu_{cs} \sim M_J^2/(E_JR')$ respectively. Also these scales minimize large logarithms in their own perturbative results. 

Resumming procedure from the factorization theorem is given as follows: We factorize $\Phi_k$ at a certain scale, i.e., the factorization scale $\mu_f$. Then each factorized part is computed at its own scale to minimize large logarithms. Finally we evolve each factorized part from its own scale to $\mu_f$ solving renormalization group (RG) equations. Since this procedure does not allow large logarithms in each factorized part, it automatically resums the large logarithms of $M_J^2/(E_J^2R'^2)$ over all through RG evolutions of $\mc{C}_k$, $J_k$, and $S_k$. 

The anomalous dimensions for the factorized parts satisfy the following RG equations: 
\be
\label{RGEs}
\frac{d}{d\ln\mu} \mc{C}_k = \gamma_{\mc{C}}^k~ \mc{C}_k, 
~~~\frac{d}{d\ln\mu} f_k (M^2) = \int^{M^2}_0 dM'^2 \gamma_f^k (M'^2) f_k (M^2-M'^2), 
\ee
where $f=J,~S$. 
From NLO results, the leading anomalous dimensions $\gamma_{\mc{C},J,S}^{(0),k}$ are written as 
\bea
\label{anomclo} 
\gamma_{\mc{C}}^{(0),k} &=& -\frac{\as C_k}{2\pi} \Bigl(2 \ln \frac{\mu^2}{p_J^{+2} t^2}+c_k\Bigr)_, \\ 
\label{anomjlo}
\gamma_{J}^{(0),k} (M^2) &=& \frac{\as C_k}{2\pi} \Biggl\{\delta(M^2) \Bigl(4\ln\frac{\mu^2}{\Lambda^2}+j_k\Bigr)-\Bigl[\frac{4}{M^2}\Bigr]_{\Lambda^2}\Biggr\}_, \\
\label{anomslo}
\gamma_{S}^{(0),k} (M^2)  &=& \frac{\as C_k}{2\pi} \Biggl\{-2 \delta(M^2)\ln\frac{\mu^2p_J^{+2}t^2}{(\Lambda^2)^2}+\Bigl[\frac{4}{M^2}\Bigr]_{\Lambda^2}\Biggr\}_,
\eea
where $c_q = j_q = 3$, $c_g = j_g = \beta_0/C_A$, and $C_k$ are $C_F$ for $k=q$ and $C_A$ for $k=g$. Also we easily check that $\Phi_k$ is scale invariant through
\be
\delta(M^2) \gamma_{\mc{C}}^{(0),k}+ \gamma_{J}^{(0),k} (M^2)  + \gamma_{S}^{(0),k} (M^2)=0.
\ee
 
In Eqs.~(\ref{anomclo})-(\ref{anomslo}) the logarithmic terms represent Sudakov logarithms. So, at leading logarithm (LL) accuracy, we resum double logarithms. Hence the result appears as $\sum_{n=0} a_k (\as L^2)^n \sim \exp(L f_0 (\as L))$, where the schematic $L$ denotes the large logarithm of $M^2/(E_J^2R'^2)$. In order to resum at next-to-leading logarithm (NLL) accuracy, 
we need the anomalous dimensions beyond LO, which can be expressed as  
\bea
\label{anomc} 
\gamma_{\mc{C}}^{k} &=& A_c \Gamma_C^{k} \ln \frac{\mu^2}{p_J^{+2} t^2}+\hat{\gamma}_c^k, \\ 
\label{anomj}
\gamma_{J}^{k} (M^2) &=& \delta(M^2) \Bigl(A_j \Gamma_C^{k} \ln \frac{\mu^2}{p_J^{+2} t^2}+\hat{\gamma}_j^k\Bigr)-\kappa_j A_j \Gamma_C^{k}\Bigl[\frac{1}{M^2}\Bigr]_{\Lambda^2}\ , \\
\label{anoms}
\gamma_{S}^{k} (M^2) &=& \delta(M^2) \Bigl(A_s \Gamma_C^{k} \ln \frac{\mu^2p_J^{+2}t^2}{(\Lambda^2)^2}+\hat{\gamma}_s^k\Bigr)-\kappa_s A_s \Gamma_C^{k}\Bigl[\frac{1}{M^2}\Bigr]_{\Lambda^2}\ ,
\eea
where $\Gamma_{C}^k = \sum_{n=0} \Gamma_{n}^k(\as/4\pi)^{n+1} $ are the cusp anomalous dimensions~\cite{Korchemsky:1987wg,Korchemskaya:1992je}, and the first two coefficients to be needed at NLL accuracy are given as
\be
\Gamma_{0}^k = 4C_k,~~~\Gamma_{1}^k = 4C_k \Biggl[\Bigl(\frac{67}{9}-\frac{\pi^2}{3}\Bigr) C_A - \frac{10}{9} n_f\Biggr]\ .
\ee
From leading anomalous dimensions in Eqs.~(\ref{anomclo}), (\ref{anomjlo}), and (\ref{anomslo}), we extract $\{A_c,A_{j},A_s, \kappa_{j},\kappa_s\} = \{-1,2,-1,1,2\}$. And the noncusp anomalous dimensions for NLL accuracy are given as $\hat{\gamma}_{c}^q=-3\as C_F/(2\pi)$, $\hat{\gamma}_{c}^g=-\as \beta_0/(2\pi)$, $\hat{\gamma}_j^k=-\hat{\gamma}_{c}^k$, and  $\hat{\gamma}_s^k=0$.

Solving RG equations in Eq.~(\ref{RGEs}) and following the method developed in Refs.~\cite{Neubert:2005nt,Becher:2006nr} we obtain the resummed result at NLL accuracy such as 
\bea 
\label{rPhik}
\bar{\Phi}_k (\rho) =  \Phi_k (M_J^2) \cdot E_J^2R'^2 &=& \exp[\mc{M}_k(\mu_c,\mu_{uc},\mu_{cs})]~\mc{C}_k(E_JR',\mu_c)  \\
&&\times\bar{J}_k \Bigl[\ln\frac{\mu_{uc}^2}{E_J^2R'^2}-\partial_{\eta}\Bigr] \bar{S}_k \Bigl[\ln\frac{\mu_{uc}^2}{E_J^2R'^2}-2\partial_{\eta}\Bigr]
\frac{e^{-\gamma_E\eta}}{\Gamma(\eta)} \rho^{-1+\eta},\nnb
\eea
where $\rho = M_J^2/(E_J^2R'^2)$, and $\bar{\Phi}_k$ is the dimensionless jet mass distribution, which can be also expressed as a ratio of $d\sigma/(dp_T^Jdyd\rho)$ over $d\sigma/(dp_T^Jdy)$. The exponentiation factor $\mc{M}_k$ at NLL accuracy is obtained as 
\bea
\label{expf}
\mc{M}_k (\mu_c,\mu_{uc},\mu_{cs}) &=& -2 S_{\Gamma}^k (\mu_{uc},\mu_{c})-2 S_{\Gamma}^k (\mu_{uc},\mu_{cs}) \\
&&+ \ln\frac{\mu_{uc}^2}{E_J^2R'^2}\Bigl(a[\Gamma_C^k](\mu_c,\mu_{uc})+a[\Gamma_C^k](\mu_{cs},\mu_{uc})\Bigr)
+a[\hat{\gamma}_j^k](\mu_c,\mu_{uc}). \nnb
\eea
Here $S_{\Gamma}^k$ and $a[f]$ are defined as 
\be
S_{\Gamma}^k (\mu_1,\mu_2) = \int^{\alpha_1}_{\alpha_2} \frac{d\as}{b(\as)} \Gamma_{C}^k(\as) \int^{\as}_{\alpha_1} \frac{d\as'}{b(\as')},~~~a[f](\mu_1,\mu_2) = \int^{\alpha_1}_{\alpha_2} \frac{d\as}{b(\as)} f(\as),
\ee
where $\alpha_{1,2} \equiv \as (\mu_{1,2})$ and $b(\as)$ is the QCD beta function given by $d\as/(d\ln\mu)$ .
In Eq.~(\ref{rPhik}) the parameter $\eta$ is defined as $\eta = 2a[\Gamma_C^k](\mu_{uc},\mu_{cs})$ and given as a positive value  since $\mu_{uc} > \mu_{cs}$. Finally $\bar{J}_k$ and $\bar{S}_k$ up to NLO are written as 
\bea
\bar{J}_q [L] &=& 1+\frac{\as C_F}{2\pi} \Bigl(\frac{7}{2}-\frac{\pi^2}{3}+\frac{3}{2}L+L^2\Bigr),\\
\bar{J}_g [L] &=& 1+\frac{\as C_A}{2\pi} \Bigl(\frac{67}{18}-\frac{5n_f}{9 C_A}-\frac{\pi^2}{3}+\frac{\beta_0}{2C_A}L+L^2\Bigr),\\
\bar{S}_k [L] &=& 1+\frac{\as C_k}{2\pi} \Bigl(-\frac{1}{2}L^2-\frac{\pi^2}{4}\Bigr).
\eea
Note that the dependence on the factorization scale does not appear in Eq.~(\ref{rPhik}) since $\Phi_k$ is scale invariant.  

One additional ingredient for the resummation at NLL accuracy is a nonglobal logarithm (NGL)~\cite{Dasgupta:2001sh,Banfi:2002hw}.
Usually NGLs appear when multiple gluons radiate near jet boundary and the gluon(s) contributes to jet observables with a limited phase space by the jet boundary. Especially large NGLs arise when there is a large energy hierarchy between in-jet and out-jet radiated gluons. 

If we consider $\Phi_k$ in the factorization theorem in Eq.~(\ref{facjmd}), $C_k~(\mc{J}_k^{-1})$ is described by the collinear mode and responsible for the out-jet radiation with a large energy. On the other hand, the collinear-soft mode for $S_k$ contributes to the jet mass through small energy radiations inside a jet. Both modes can read the jet boundary and there is a large energy difference between them. 
Therefore we expect nonneglible NGLs contributions to the jet mass distribution. The mechanism of generating NGLs here is similar with the hemisphere jet mass distribution~\cite{Dasgupta:2001sh,Becher:2016omr}. Actually it has been found that the resummed result of leading NGLs for a narrow jet has the same form as one for the hemisphere jet mass~\cite{Banfi:2010pa,Dasgupta:2012hg}. The difference would be the scale choices relevant to two modes that read a jet boundary and have a large energy difference.  

In order to guess the size of the NGL contribution to the jet mass, we use the resummed result of leading NGLs for the hemisphere jet mass distribution in the large $N_c$ limit~\cite{Dasgupta:2001sh}. Here the leading NGLs appear from two loop, and the resummed result is schematically given as $\sum_{n=2} a_{\rm{NG}}^n (\as L)^n$ and contributes at NLL accuracy. From the resummed result in Ref.~\cite{Dasgupta:2001sh}, we estimate NGL contributions to the jet mass at NLL accuracy such as   
\be\label{RNGL}
\Delta^{k=q,g}_{\mr{NG}} (\mu_{c},\mu_{cs}) = \exp\Biggl(-C_A C_k \frac{\pi^2}{3} \Bigl(\frac{1+(at)^2}{1+(bt)^c}\Bigr) t^2 \Biggr)\ ,
\ee
where 
\be
t=\frac{1}{\beta_0} \ln{\frac{\as(\mu_{cs})}{\as(\mu_{c})}} \sim -\frac{1}{\beta_0}\ln \Bigl(1-\frac{\beta_0}{4\pi} \as(\mu_c) \ln \frac{\mu_c^2}{\mu_{cs}^2}\Bigr)\ .
\ee 
Here the fit parameters from the parton-shower are given by $a=0.85 C_A,~b=0.86C_A$, and $c=1.33$~\cite{Dasgupta:2001sh}.
If we put $\mu_c = E_JR'=p_T^JR$ and $\mu_{cs} = M_J^2/(p_T^JR)$ into Eq.~(\ref{RNGL}), the result coincides with the resummed results of the NGLs in the small $R$ limit in Ref.~\cite{Dasgupta:2012hg}

\begin{figure*}[t]
\begin{center}
\includegraphics[width=16cm]{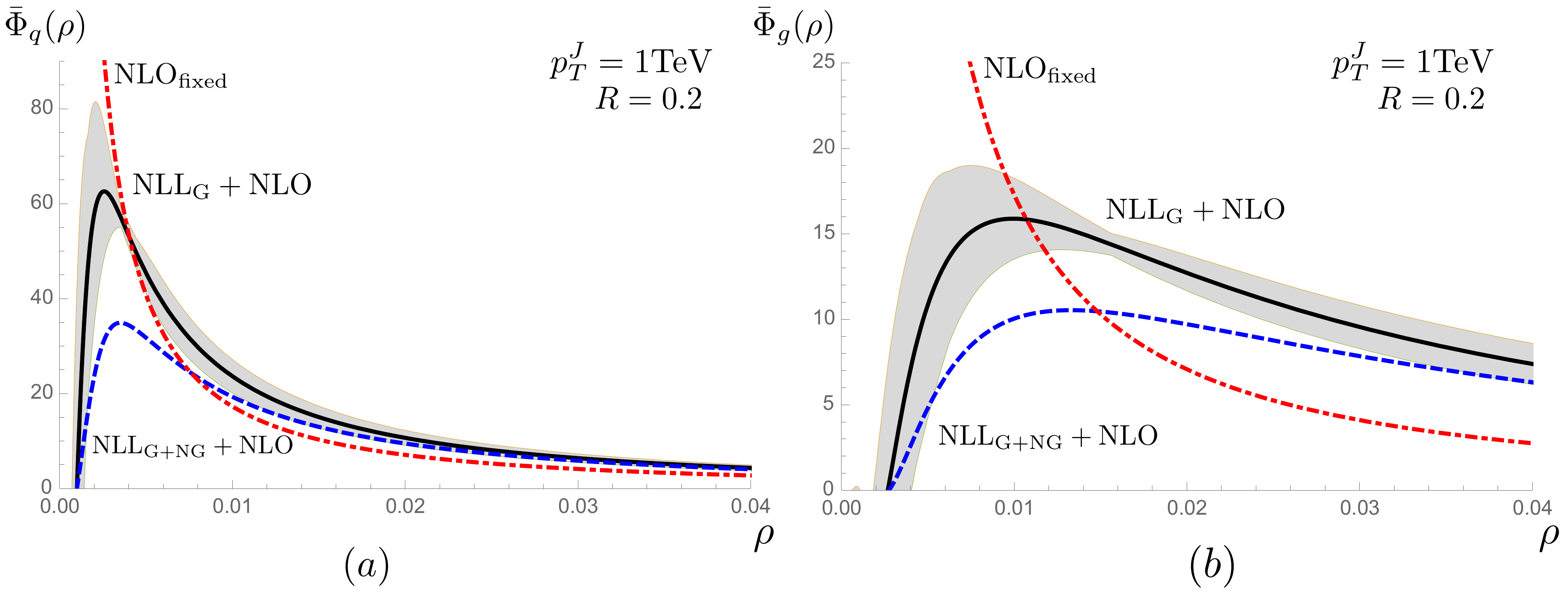}
\end{center}
\vspace{-0.6cm}
\caption{\label{fig1} \baselineskip 3.0ex 
Jet mass distributions for a quark jet (FIG.~\ref{fig1}-(a)) and a gluon jet (FIG.~\ref{fig1}-(b)) with $p_T^J =1~\mr{Tev}$ and $R=0.2$. Here $\rho = M_J^2/(p_T^JR)^2$. The thick lines ($\mr{NLL_{G}+NLO}$) denote NLL resummed results with the fixed NLO corrections and the dashed lines ($\mr{NLL_{G+NG}+NLO}$) denote the resummed results with NGLs. The dot-dashed lines ($\mr{NLO_{fixed}}$) are the results at the fixed NLO. Gray bands for $\mr{NLL_{G}+NLO}$ represent scale variations from $\mu_{c,uc,cs} = 2\mu_{c,uc,cs}^{0}$ to $\mu_{c,uc,cs}^{0}/2$. 
} 
\end{figure*}

\begin{figure*}[t]
\begin{center}
\includegraphics[width=16cm]{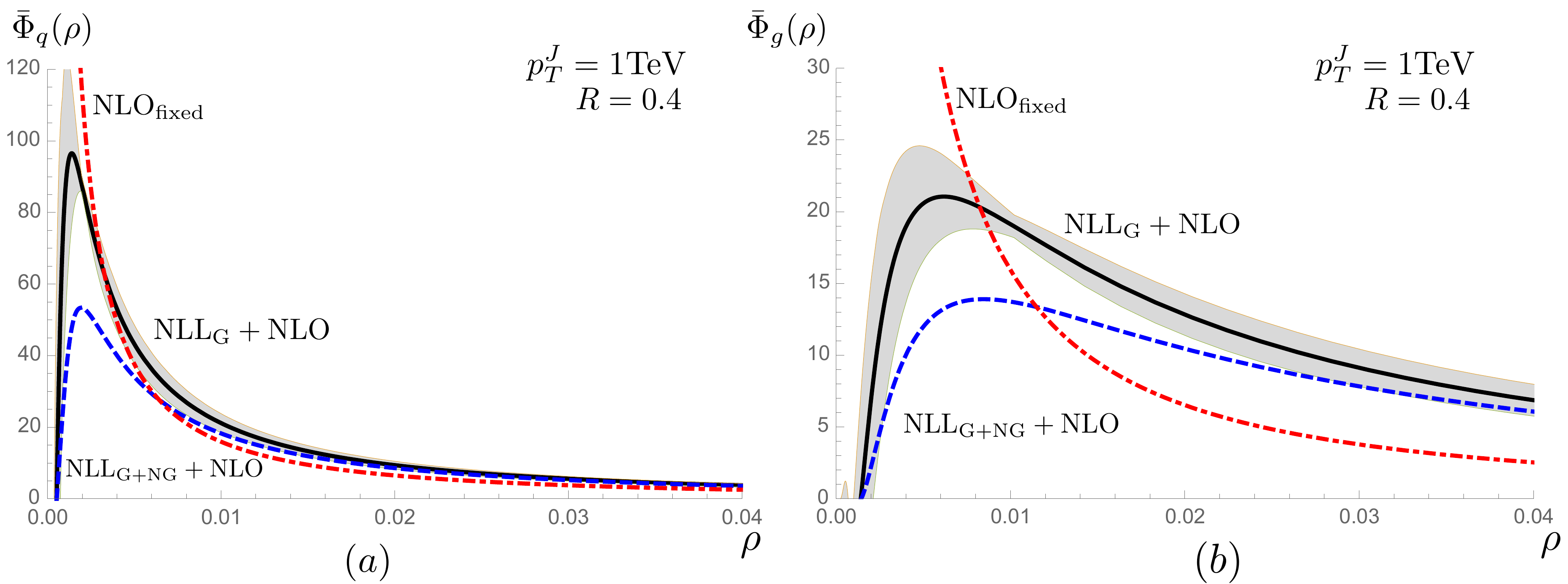}
\end{center}
\vspace{-0.6cm}
\caption{\label{fig2} \baselineskip 3.0ex 
Jet mass distributions for a quark jet (FIG.~\ref{fig2}-(a)) and a gluon jet (FIG.~\ref{fig2}-(b)) with $p_T^J =1~\mr{Tev}$ and $R=0.4$. Here the expressions are the same as FIG.~\ref{fig1}. } 
\end{figure*}

Using Eq.~(\ref{rPhik}), we show NLL resummed results plus the fixed NLO corrections (NLL+NLO) for $(p_T^J,R)=(1~\mr{TeV},0.2)$ and $(p_T^J,R)=(1~\mr{TeV},0.4)$ in FIGs.~\ref{fig1} and \ref{fig2} respectively. 
The default collinear, ultracollinear, and collinear-soft scales have been chosen as $(\mu_{c}^0,\mu_{uc}^0,\mu_{cs}^0) = (p_T^JR,M_J,M_J^2/(p_T^J R)) = p_T^J R(1,\sqrt{\rho},\rho)$, where $\rho = M_J^2/(p_T^JR)^2$. In order to avoid Landau pole as $\rho$ goes to zero, we introduce a fixed value, $\rho_0$, near the zero point. Then, for the region $\rho <\rho_0$, we make the collinear-soft scale finally freeze as some value, which is slightly above the Landau pole~\cite{Dai:2017dpc,Beneke:2011mq}. In order to implement it we introduce the collinear-soft scale profile such as 
\be
\label{csoftpf} 
\mu_{cs}^{pf} = 
\left\{
\begin{array}{rl} 
& p_T^J R\rho~~~\mr{if}~\rho \ge \rho_0,  \\ 
& \mu_{min} + a  p_T^J R \rho^2~~~\mr{if}~\rho < \rho_0,   \end{array} \right.
\ee
where we set $\mu_{min} = 0.3~\mr{GeV}$. And $\rho_0$ and $a$ are chosen in order that $\mu_{cs}^{pf}$ be smoothly continuous at $\rho_0$. Accordingly the collinear scale is given as $\mu_{uc}^{pf} = \sqrt{p_T^J R\cdot\mu_{cs}^{pf}}$.  

As seen from FIGs.~\ref{fig1} and \ref{fig2}, the results at the fixed NLO in $\as$ that are shown in Eqs.~(\ref{Phiq}) and (\ref{Phig}) diverge in the small jet mass region. However, the NLL resummed distributions show more reliable results reflecting Sudakov suppression. While the distribution of the jet initiated by a quark has a sharp shape near the zero point, the jet initiated by a gluon has a broader shape and the peak shifts positively due to a large color factor $C_A$. We have also included the resummed results of leading NGLs shown in Eq.~(\ref{RNGL}), which give significant suppressions around the peak regions. 
In FIGs.~\ref{fig1} and \ref{fig2}, we have not considered nonperturbative effects, that would be examined by comparison with experimental data in the future. 

Even though the jet mass distributions are independent of the factorization scale, we have some arbitrariness to choose $\mu_{c,uc,cs}$ for each factorized part. In order to see the scale dependence we vary the scales from $\mu_{c,uc,cs} = 2\mu_{c,uc,cs}^{0}$ to $\mu_{c,uc,cs}^{0}/2$, then obtain scale variations for the results at $\mr{NLL_G+NLO}$, where the subscript `G' represents the global logarithms without NGLs. The variances are expressed as gray bands in FIGs.~\ref{fig1} and \ref{fig2}. We checked that the errors have been significantly reduced after including the fixed NLO corrections in the NLL resummed results.

\section{Conclusion}

We present a factorization theorem for the jet mass distribution with a given $p_T^J$ (or $E_J$) for the inclusive jet process. As a result the scattering cross section is factorized as partonic scattering cross sections, the FFJs, and the jet mass distributions $\Phi_k (M_J^2)$ up to NLO in $\as$.  
%$h_1 h_2 \to JX$. Our results can be also applied to deep inelastic scattering, $e^+e^-$ annihilation and the single jet production:  $h_1h_2\to J+L$, where $L$ is a color singlet particle.
Since the decoupled $\Phi_k$ can be obtained from the ratio of two cross sections such as $d\sigma/(dp_T^JdM_J^2)$ and $d\sigma/dp_T^J$, it is given to be scale invariant. Also $\Phi_k$ can be calculated independently of a hard scattering process once a jet with high $p_T^J$ (or $E_J$) and small $R$ is given. So it promises a consistent treatment for examining QCD jet mass distributions in various processes. 
%More precise estimations of $F_{q,g}$, while including the higher order perturbative results and nonperturbative effects, are needed to clarify QCD jet substructures. 

In the region $M_J \ll E_JR'$, the jet mass distributions can be additionally factorized as the collinear, ultracollinear, and collinear-soft parts. Using the factorized result in SCET we have consistently resummed large logarithms of $M_J/(E_JR')$ up to NLL accuracy. 
For more precise results we need the higher order results of the collinear-soft functions $S_k$ or the integrated jet functions $\mc{J}_k$ in the factorization framework. 
Especially the integrated jet functions beyond one loop would be very useful not only for the jet mass but also for various other jet substructures. Also we need to understand NGLs and clustering logarithms~\cite{Banfi:2005gj,Delenda:2006nf} at deeper level.  
Finally we note that the factorization theorem here can give a firm basis to consider the groomed jet mass distributions~\cite{Dasgupta:2013ihk,Larkoski:2014wba,Frye:2016aiz} for inclusive processes since most subprocesses are similar except that the groomed observable has more restrictions inside a jet.

\acknowledgments

C.~Kim is grateful to Junegone Chay for helpful comments. 
C.~Kim was supported by Basic Science Research Program through the National Research Foundation of Korea (NRF) funded by the Ministry of Science and ICT (Grants No.~NRF-2014R1A2A1A11052687, No.~NRF-2017R1A2B4010511).

%%%%%%%%%%%%%%%%%%%%%%%%%%%%%%%%%%%%%%%%%%%%%%%%%%%%%%%%%%%%%%%%%%%%%%
%%%%%%%%%%%%%%%%%%%%%%%%%%%%% Bibliography %%%%%%%%%%%%%%%%%%%%%%%%%%%
%%%%%%%%%%%%%%%%%%%%%%%%%%%%%%%%%%%%%%%%%%%%%%%%%%%%%%%%%%%%%%%%%%%%%%

%%%%%%%%%%%%%%%%%%%%%%%%%%%%%%%%%%%%%%%%%%%%%%%%%%%%%%%%%%%%%%%%%%%%%%

\end{document}